\documentclass{article}
\usepackage{arxiv}

\usepackage[utf8]{inputenc} 
\usepackage[T1]{fontenc}    

\usepackage{authblk}
\usepackage[hyperindex,breaklinks,colorlinks,urlcolor=blue,citecolor=blue,linkcolor=blue]{hyperref}
\usepackage{url}            
\usepackage{booktabs}       
\usepackage{amsfonts}       
\usepackage{nicefrac}       
\usepackage{microtype}      
\usepackage{graphicx}
\usepackage{natbib}

\usepackage{doi}

\newcommand{\UMDphy}{Department of Physics, University of Maryland, College Park, Maryland 20742, USA}
\newcommand{\QTC}{Quantum Technology Center, University of Maryland, College Park, Maryland 20742, USA}
\newcommand{\UMDEECS}{Department of Electrical and Computer Engineering, University of Maryland, College Park, Maryland 20742, USA}

\renewcommand{\headeright}{}
\renewcommand{\undertitle}{\href{https://github.com/hakalasonja/tempo}{\texttt{https://github.com/hakalasonja/tempo}}}
\renewcommand{\shorttitle}{\url{https://github.com/hakalasonja/TEMPO.git}}

\begin{document}

\title{TEMPO: A Python Package for Time Evolution of Pulse Sequences in QuTiP}
\date{\vspace{-0.2cm}}

\author[1, 2]{Jner Tzern Oon$^*$}
\author[1]{Sonja A. Hakala$^*$}
\author[1]{George A. Witt}
\author[1, 2, 3]{Ronald Walsworth}

\affil[1]{\UMDphy}
\affil[2]{\QTC}
\affil[3]{\UMDEECS}

\def\thefootnote{*}\footnotetext{These authors contributed equally to this work}

\maketitle

\section{Summary}\label{summary}

TEMPO (Time-dependent Evolution of Multiple Pulse Operations) offers
accessible and efficient simulations of pulse sequences in Python, using
the suite of master equation solvers available in the Quantum Toolbox in
Python (QuTiP). It enables straightforward definition of pulse sequence
structures, including any underlying time-dependent Hamiltonians and
pulse timing information, and faster simulations of pulse sequence
dynamics (compared to naive implementations using QuTiP) while remaining
compatible with the existing collection of QuTiP subpackages. Given the
ubiquitous use of pulse sequences throughout quantum
information/computing sciences, magnetic resonance studies, and quantum
metrology, this work has immediate relevance to a wide array of research
applications.

\section{Statement of Need}\label{statement-of-need}

Pulse sequences typically contain a series of discrete operations
(pulses) using radio frequency, microwave, or optical fields. Their
application for quantum control has an extensive history across atomic
physics (\cite{Vitanov2001}); nuclear
magnetic resonance (\cite{Levitt2013});
solid-state spin systems used in quantum technologies
(\cite{Barry2020}); and a broad range of
other platforms. In recent years, research into quantum technologies has
driven the development of advanced software tools for numerical
simulations of quantum systems (\cite{Fingerhuth2018}). In particular, the QuTiP (Quantum Toolbox in Python)
framework provides open-source tools for simulations of open quantum
systems, and has received prolific use across numerous quantum
applications (\cite{Johansson2012, Johansson2013}),
solvers that are native to QuTiP, TEMPO provides two key advantages for
numerical simulations of pulse sequence dynamics.

\textbf{Ease of Use:} By incorporating both the characteristics of a
Hamiltonian and its time constraints as necessary properties, pulses are
first constructed individually, then collated to form a `pulse
sequence'. Time evolution is performed without the need to manually
deactivate each pulse Hamiltonian outside its given time interval. Using
pulse `recipes' in TEMPO, the creation of pulses with overlapping
functional forms is streamlined, with parameters that can be tuned for
individual pulses.

\textbf{Faster Executions of Time Evolution:} TEMPO organizes each pulse
sequence as a series of time segments, preserving only the pulses that
are active within each segment. This avoids overheads incurred by
repeated inspections of inactive pulse(s), significantly speeding up
evaluation of system evolution.

\begin{figure}
\centering
\includegraphics[width=\linewidth]{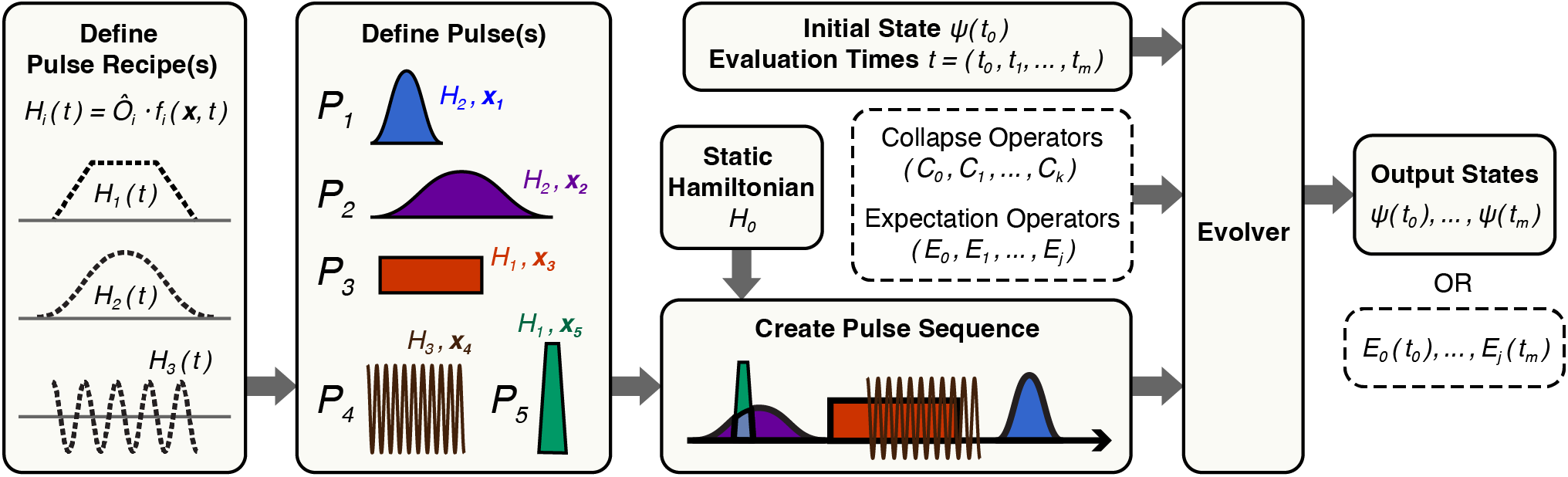}
\caption{Outline of steps in TEMPO used to perform time evolution due to
a pulse sequence. First, each pulse recipe contains the functional form
of a time-dependent Hamiltonian
\(H_i(t) = \hat{O}_i \cdot f_i(\mathbf{x}, t)\). Pulse recipe(s) are
used to create individual pulse(s) \(P_1, \dots, P_n\), with individual
parameter settings \(\mathbf{x_1}, \dots, \mathbf{x_n}\), respectively.
The pulses are organized into a sequence along with an optional static
Hamiltonian. Next, the pulse sequence is provided to the Evolver, with
an initial system state \(\psi(t_0)\) and an array of evaluation times
\((t_0, t_1, \dots, t_m)\). Time evolution returns the state
\(\psi(t_0), \dots, \psi(t_m)\) at these times; if operators
\(E_0, \dots, E_j\) are provided, the operator expectation values
\(E_0(t_0), \dots, E_j(t_0)\) are calculated instead. \label{fig:1}}
\end{figure}

\section{Usage}\label{usage}

There are five main classes that make up a simulation in TEMPO:
\texttt{Pulse\_Recipe}, \texttt{Pulse}, \texttt{Hamiltonian},
\texttt{Pulse\_Sequence}, and \texttt{Evolver}. In general, simulations
can be easily executed by following the steps outlined below, which are
illustrated in \autoref{fig:1}.

\begin{enumerate}
\def\labelenumi{\arabic{enumi}.}
\item
  Create a pulse recipe by defining the functional form of the
  time-dependent Hamiltonian
  \[\hat H_i(t) = \hat O_i \cdot f_i(\mathbf{x},  t).\] The user
  provides an operator/matrix \(\hat O\) and scalar function
  \(f(\mathbf{x}, t)\) that depends on input parameters \(\mathbf{x}\)
  and time \(t\).
\item
  Create individual pulse(s) \(P_1, \dots, P_n\) by providing a pulse
  recipe, individual parameter settings
  \(\mathbf{x_1}, \dots, \mathbf{x_n}\), and pulse timing information.
\item
  Generate a pulse sequence by inputting the pulses \(P_1, \dots, P_n\),
  and optionally a time-independent (static) Hamiltonian.
\item
  To evolve the system in time, the pulse sequence is provided to the
  Evolver class along with the initial system state \(\psi(t_0)\) (state
  vector or density matrix) and an array of time points
  \(t_0, t_1, \dots, t_m\). The system state is returned at these times.
  It is also possible to provide collapse operators for evolution using
  the Linbladian (open) master equation, along with operators for
  calculation of expectation values at these times.
\item
  The Evolver returns the system state \(\psi(t_0), \dots, \psi(t_m)\)
  or operator expectation values at the times provided.
\end{enumerate}

\begin{figure}
\centering
\includegraphics[width=\linewidth]{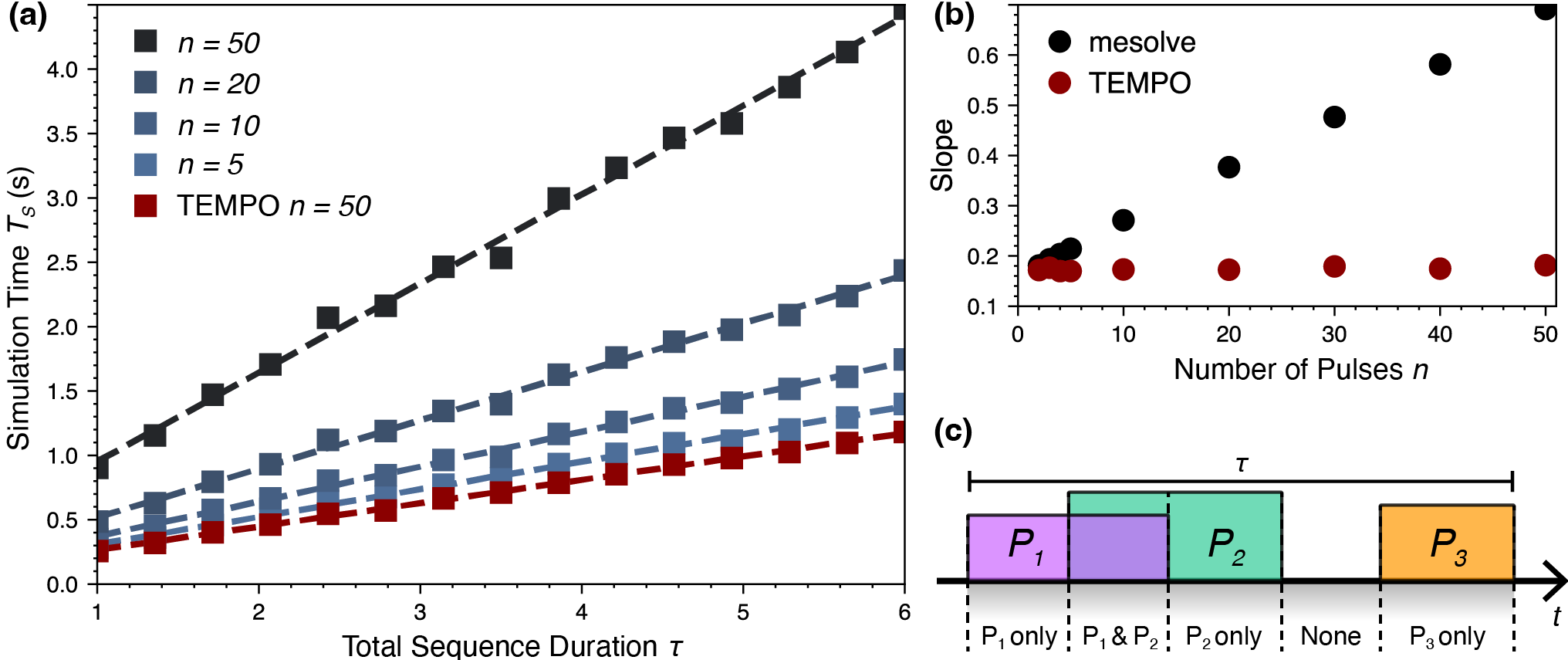}
\caption{(a) Average simulation wall-clock time \(T_S\) as a function of
pulse sequence duration \(\tau\) without using TEMPO, for number of
pulses \(n=5,10,20,50\), shown in shades of blue. Average simulation
times using TEMPO are shown in red at high pulse number count
\((n=50)\). Best-fit curves are displayed as dashed lines, with slope
values plotted in (b) for measurements with TEMPO in red and without
TEMPO (via QuTiP's mesolve function) in black. (c) Example visualization
of pulse sequence segmentation performed by TEMPO, where \(n=3\) pulses
result in \(2n-1=5\) time segments for solver evaluation. All
measurements of \(T_S\) are average values over 20 repeated runs.
\label{fig:2}}
\end{figure}

\section{Efficient Simulation
Times}\label{efficient-simulation-times}

Using the existing master equation solvers available in QuTiP, TEMPO
provides significant speedups in simulations of multi-pulse sequences.
Without TEMPO, the simulation (wall-clock) time \(T_S\) generally follow
a linear trend with respect to 1) the total duration of a pulse sequence
\(\tau\), and 2) the number of pulses \(n\), roughly obeying
$$T_S \propto n \cdot \tau.$$
This dependence is highlighted in
\autoref{fig:2}(a), which shows measurements of the average simulation
time with varying \(\tau\). Each set of measurements, shown as points of
the same color, is performed for a fixed number of pulses \(n\). Without
TEMPO (shades of blue), the fitted slope for each set of measurements
increases with \(n\), exhibiting a linear dependence as shown in
\autoref{fig:2}(b). In contrast, the use of TEMPO (red) results in
simulation times that are nearly independent of the number of pulses
\(n\), instead following \(T_S \propto \tau\). As a result, the slope of
each best-fit line using TEMPO stays roughly constant with respect to
\(n\), as shown in \autoref{fig:2}(b).

This advantage can be understood by considering how the solvers are
constructed natively (without TEMPO). At each timestep during the pulse
sequence, the solver typically inspects if each pulse is active at this
time. This results in \(n\) checks at each timestep and thus the rough
linear dependence of \(T_S\) with respect to \(n\).

Instead, TEMPO first divides the pulse sequence into consecutive time
segments for efficient time evolution, illustrated by an example in
\autoref{fig:2}(c). By creating segment breaks at the start and end of
each pulse, TEMPO preserves only the active pulses in each of the
\(2n-1\) segments. The solver is then executed consecutively across each
segment, without any redundant checks of inactive pulses. As a result,
simulation times are largely independent of \(n\), besides minor
overheads from repeated use of the solver.

\section{Acknowledgements}\label{acknowledgements}

We thank Saipriya Satyajit, Katrijn Everaert, Declan Daly, Kevin Olsson
and John Blanchard for testing the package and providing feedback during
development. This work is supported by, or in part by, the DEVCOM Army
Research Laboratory under Contract Numbers W911NF1920181 and
W911NF2420143; the DEVCOM ARL Army Research Office under Grant Number
W911NF2120110; the U.S. Air Force Office of Scientific Research under
Grant Number FA95502210312; and the University of Maryland Quantum
Technology Center.

\bibliographystyle{apalike}
\bibliography{paper.bib} 

\begin{thebibliography}{}

\bibitem[Barry et~al., 2020]{Barry2020}
Barry, J.~F., Schloss, J.~M., Bauch, E., Turner, M.~J., Hart, C.~A., Pham, L.~M., and Walsworth, R.~L. (2020).
\newblock Sensitivity optimization for nv-diamond magnetometry.
\newblock {\em Reviews of Modern Physics}, 92(1).

\bibitem[Fingerhuth et~al., 2018]{Fingerhuth2018}
Fingerhuth, M., Babej, T., and Wittek, P. (2018).
\newblock Open source software in quantum computing.
\newblock {\em PLOS ONE}, 13(12):e0208561.

\bibitem[Johansson et~al., 2012]{Johansson2012}
Johansson, J., Nation, P., and Nori, F. (2012).
\newblock Qutip: An open-source python framework for the dynamics of open quantum systems.
\newblock {\em Computer Physics Communications}, 183(8):1760–1772.

\bibitem[Johansson et~al., 2013]{Johansson2013}
Johansson, J., Nation, P., and Nori, F. (2013).
\newblock Qutip 2: A python framework for the dynamics of open quantum systems.
\newblock {\em Computer Physics Communications}, 184(4):1234–1240.

\bibitem[Levitt, 2013]{Levitt2013}
Levitt, M.~H. (2013).
\newblock Spin dynamics.

\bibitem[Vitanov et~al., 2001]{Vitanov2001}
Vitanov, N., Fleischhauer, M., Shore, B., and Bergmann, K. (2001).
\newblock Coherent manipulation of atoms molecules by sequential laser pulses.

\end{thebibliography}

\end{document}